\documentclass[usenatbib,fleqn]{mnras}

\usepackage{newtxtext,newtxmath}
\usepackage[T1]{fontenc}
\usepackage{ae,aecompl}

\usepackage{graphicx}	% Including figure files
\usepackage{amsmath}	% Advanced maths commands
\usepackage{amssymb}	% Extra maths symbols

\usepackage{times}
\newcommand{\wmap}{\textsl{WMAP}}
\newcommand{\planck}{\textsl{Planck}}
\newcommand{\rosat}{\textsl{ROSAT}}
\newcommand{\suzaku}{\textsl{SUZAKU}}
\newcommand{\beq}{\begin{equation}}
\newcommand{\eeq}{\end{equation}}
\newcommand{\beqa}{\begin{eqnarray}}
\newcommand{\eeqa}{\end{eqnarray}}

\title[A Search for Gas Filaments]{A Search for Warm/Hot Gas Filaments Between Pairs of SDSS Luminous Red Galaxies}

\author[Tanimura et al.]{Hideki Tanimura,$^{1,2}$\thanks{E-mail: hideki.tanimura@ias.u-psud.fr}
Gary Hinshaw,$^{2,3,4}$
Ian G. McCarthy,$^{5}$ 
Ludovic Van Waerbeke,$^{2,3}$\newauthor
Nabila Aghanim,$^{1}$
Yin-Zhe Ma,$^{6,7}$
Alexander Mead,$^{2}$
Alireza Hojjati$^{2}$
and Tilman Tr\"{o}ster$^{2}$
\\
$^{1}$Institut d'Astrophysique Spatiale, CNRS (UMR 8617), Universit\'{e} Paris-Sud, B\^{a}timent 121, Orsay, France\\
$^{2}$Department of Physics and Astronomy, University of British Columbia, Vancouver, BC V6T 1Z1, Canada\\
$^{3}$Canadian Institute for Advanced Research, 180 Dundas St W, Toronto, ON M5G 1Z8, Canada\\
$^{4}$Canada Research Chair in Observational Cosmology\\
$^{5}$CAstrophysics Research Institute, Liverpool John Moores University, Liverpool, L3 5RF, United Kingdom\\
$^{6}$Astrophysics and Cosmology Research Unit, School of Chemistry and Physics,
University of KwaZulu-Natal, Durban, South Africa\\
$^{7}$NAOC-UKZN Computational Astrophysics Centre (NUCAC), University of KwaZulu-Natal, Durban, 4000
}

\pubyear{2017}

\begin{document}
\label{firstpage}
\pagerange{\pageref{firstpage}--\pageref{lastpage}}
\maketitle

\begin{abstract}
We search the \planck\ data for a thermal Sunyaev-Zel'dovich (tSZ) signal due to gas filaments between pairs of Luminous Red Galaxies (LRG's) taken from the Sloan Digital Sky Survey Data Release 12 (SDSS/DR12).  We identify $\sim$260,000 LRG pairs in the DR12 catalog that lie within 6--10 $h^{-1} \mathrm{Mpc}$ of each other in tangential direction and within 6 $h^{-1} \mathrm{Mpc}$ in radial direction.  We stack pairs by rotating and scaling the angular positions of each LRG so they lie on a common reference frame, then we subtract a circularly symmetric halo from each member of the pair to search for a residual signal between the pair members.  We find a statistically significant (5.3$\sigma$) signal between LRG pairs in the stacked data with a magnitude $\Delta y = (1.31 \pm 0.25) \times 10^{-8}$.  The uncertainty is estimated from two Monte Carlo null tests which also establish the reliability of our analysis. Assuming a simple, isothermal, cylindrical filament model of electron over-density with a radial density profile proportional to $r_c/r$ (as determined from simulations), where $r$ is the perpendicular distance from the cylinder axis and $r_c$ is the core radius of the density profile, we constrain the product of over-density and filament temperature to be $\delta_c \times (T_{\rm e}/10^7 \, {\rm K}) \times (r_c/0.5h^{-1} \, {\rm Mpc}) = 2.7 \pm 0.5$.  To our knowledge, this is the first detection of filamentary gas at over-densities typical of cosmological large-scale structure. We compare our result to the BAHAMAS suite of cosmological hydrodynamic simulations \citep{McCarthy2017} and find a slightly lower, but marginally consistent Comptonization excess, $\Delta y = (0.84 \pm 0.24) \times 10^{-8}$. 
\end{abstract}

\begin{keywords}
galaxies -- groups -- clusters -- halos -- filaments -- large-scale structure -- cosmology
\end{keywords}

%%%%%%%%%%%%%%%%%%%%%%%%%%%%%%%%%%%%%%%%%%%%%%%%%%

%%%%%%%%%%%%%%%%% BODY OF PAPER %%%%%%%%%%%%%%%%%%

\section{Introduction}
In the now-standard $\Lambda$CDM cosmology, more than $\sim$95\% of the energy density in the universe is in the form of dark matter and dark energy, whereas baryonic matter only comprises $\sim$5\% \citep{Planck2016-I,Hinshaw2013}. 

At high redshift ($z \gtrsim 2$), most of the expected baryons are found in the Ly$\alpha$ absorption forest: the diffuse, photo-ionized intergalactic medium (IGM) with a temperature of $10^4$--$10^5$ K (e.g., \citealt{Weinberg1997, Rauch1997}).   However, at redshifts $z \lesssim 2$, the observed baryons in stars, the cold interstellar medium, residual Ly$\alpha$ forest gas, OVI and BLA absorbers, and hot gas in clusters of galaxies account for only $\sim$50\% of the expected baryons -- the remainder has yet to be identified (e.g., \citealt{Fukugita2004, Nicastro2008, Shull2012}). Hydrodynamical simulations suggest that 40--50\% of baryons could be in the form of shock-heated gas in a cosmic web between clusters of galaxies.  This so-called Warm Hot Intergalactic Medium (WHIM) has a temperature range of $10^5$--$10^7$ K \citep{Cen2006}.  The WHIM is difficult to observe due to its low density: several detections in the far-UV and X-ray have been reported, but none are considered definitive \citep{Yao2012}.

The thermal Sunyaev-Zel'dovich (tSZ) effect \citep{Zeldovich1969, Sunyaev1970, Sunyaev1972, Sunyaev1980} arises from the Compton scattering of CMB photons as they pass through hot ionized gas along the line of sight.  The signal provides an excellent tool for probing baryonic gas at low and intermediate redshifts.  \cite{Atrio2006} and \cite{Atrio2008} suggest that electron pressure in the WHIM would be sufficient to generate potentially observable tSZ signals.  However, the measurement is challenging due to the morphology of the source and the relative weakness of the signal.

The \planck\ satellite mission has produced a full-sky tSZ map with 10 arcmin angular resolution.  In addition to the numerous galaxy clusters detected in tSZ by \planck\ , a significant tSZ signal in the inter-cluster region between the merging clusters of A399$-$A401 is reported in \cite{Planck2013IR-VIII} and \cite{Bonjean2018}.  In conjunction with \rosat\ X-ray data, they estimate the temperature and density of the inter-cluster gas.

On a larger scale, luminous red galaxies (LRG's) are powerful tracers of large-scale structure of the universe.  These early-type, massive galaxies, selected on the basis of color and magnitude, consist mainly of old stars with little ongoing star formation.  LRG's typically reside in the centers of galaxy groups and clusters and have been used to detect and characterize the remnants of baryon acoustic oscillations (BAO) at low to intermediate redshift \citep{Eisenstein2005, Kazin2010, Anderson2014}. 

\cite{Clampitt2016} searched for evidence of massive filaments between proximate pairs of LRG's taken from the Sloan Digital Sky Survey seventh data release (SDSS DR7).   Using weak gravitational lensing signal, stacked on 135,000 pairs of LRG's, they find evidence for filament mass at $\sim$4.5$\sigma$ confidence. Similarly, \cite{Epps2017} detect the weak lensing signal using the Canada France Hawaii Telescope Lensing Survey (CFHTLenS) mass map from stacked filaments between SDSS-III/BOSS LRG's at 5$\sigma$ confidence.

\cite{Waerbeke2014}, \cite{Ma2015} and \cite{Hojjati2015} report correlations between gravitational lensing and tSZ signals in the field using the CFHTLenS mass map and \planck\ tSZ map.  Similarly, \cite{Hill2014} reports a statistically significant correlation between the \planck\ CMB lensing potential and the \planck\ tSZ map.  These results show clear evidence for diffuse gas tracing dark matter.  Further, in the context of a halo model, there is clear evidence for contributions from both the one- and two-halo terms, but there is no statistically significant evidence for contributions from diffuse, unbound gas not associated with (correlated) collapsed halos.

In this paper, we use the \planck\ tSZ map from the 2015 data release \citep{Planck2016-I} to search for evidence of hot gas in filaments between proximate pairs of LRG's taken from the Sloan Digital Sky Survey twelfth data release  \citep{Alam2015}: SDSS DR12 LRG \citep{Prakash2016}.  Since the signal-to-noise ratio of the \planck\ $y$ map is not high enough to trace individual filaments on this scale, we employ a stacking method to search for an average signal between many pairs.  We find a mean filament signal of $\Delta y = (1.31 \pm 0.25) \times 10^{-8}$ with 5.3$\sigma$ significance.  We compare the filament signal with predictions from the BAHAMAS suite of hydrodynamic simulations \citep{McCarthy2017} and find marginally consistent results for two cosmologies (\wmap9 and \planck\ 2013) in the suite.  Throughout this work, we adopt a $\Lambda$CDM cosmology with $\Omega_{\rm m} = 0.3$, $\Omega_{\Lambda} = 0.7$, and $H_0 = 70$ km s$^{-1}$ Mpc$^{-1}$ for conversion of redshifts into distances.

%%%%%%%%%%%%%%%%%%%%%%%%%%%%%%%%%%%%%%%%%%%%%%%%%%%%%%%%%%%%%
%
%                   Data
%
%%%%%%%%%%%%%%%%%%%%%%%%%%%%%%%%%%%%%%%%%%%%%%%%%%%%%%%%%%%%%

\section{Data}
We use three data sets in this analysis: 1) the Luminous Red Galaxy catalog from the Sloan Digital Sky Survey twelfth data release\footnote{http://www.sdss.org/dr12/spectro/spectro\_access} (SDSS DR12 LRG, N$\sim$1,400,000, \cite{Prakash2016}), 2) the \planck\ Comptonization $y$ map\footnote{http://pla.esac.esa.int/pla/\#results} from the 2015 data release \citep{Planck2016-I}, and 3) the BAHAMAS suite of cosmological hydrodynamic simulations \citep{McCarthy2017}.  We describe each briefly, in turn.

\subsection{LRG pair catalog}
\label{sec:lrg}
SDSS Data Release 12 (DR12) is the final release of data from SDSS-III.  The catalog provides the position, spectroscopic redshift, and classification type for each object.  We extract objects identified as {\tt sourcetype=LRG}.  The stellar masses of these objects have been estimated by three different groups\footnote{http://www.sdss.org/dr12/spectro/galaxy}.  We use the estimate based on a principal component analysis method by \cite{Chen2012}, which uses stellar evolution synthesis models from \cite{Bruzual2003}, and a stellar initial mass function from \cite{Kroupa2001}.  For the stacking analysis, we select LRG's with $M_{\ast} > 10^{11.3} M_{\odot}$.  According to the $M_{\ast}$--$Y_{500}$ scaling relation reported in \cite{Planck2013IR-XI} ($Y_{500}$ is the Comptonization parameter integrated over a sphere of radius $R_{ 500}$), these LRG's should have a central tSZ signal-to-noise ratio of order unity.  Since our analysis requires us to estimate and subtract the tSZ signal associated with the halos of the individual LRG's, this cut enhances the reliability of that estimation, via the procedure described in \S\ref{sec:halo}.

Not all LRG's are central galaxies in massive halos.  \cite{Hoshino2015} find that, at a halo mass of $10^{14.5} M_{\odot}$, only 73\% of LRG's are central, lower than the previous estimate of 89 \% obtained from correlation studies \citep{Reid2009}.  To minimize the fraction of satellite LRG's in our sample (which could bias a filament signal) we select the locally most-massive LRG's (based on stellar mass) using a criterion that is analogous to that used in \cite{Planck2013IR-XI}: we reject a given galaxy if a more massive galaxy resides within a tangential distance of 1.0 $h^{-1} \mathrm{Mpc}$ and within a radial velocity difference of $|c \Delta {\it z}| < 1000$ km s$^{-1}$.

\begin{figure*}
\begin{center}
\begin{minipage}{0.4\linewidth}
\includegraphics[width=\linewidth]{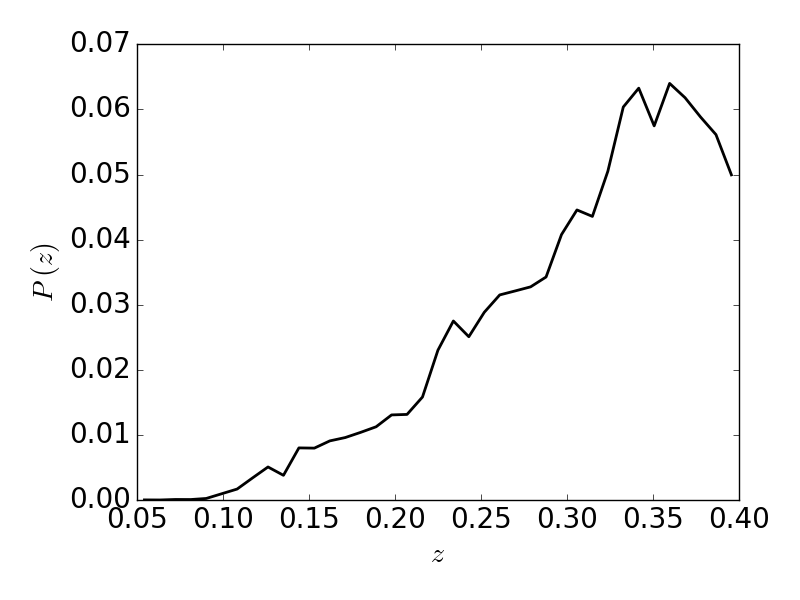}
\end{minipage}
\begin{minipage}{0.4\linewidth}
\includegraphics[width=\linewidth]{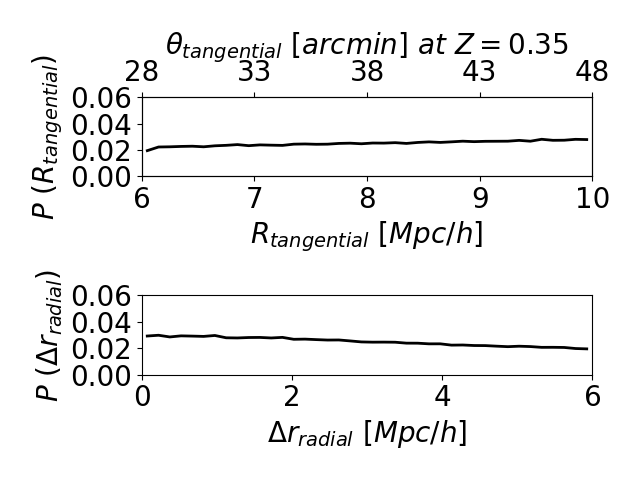}
\end{minipage}
\caption{{\it Left}: The redshift distribution of LRG pairs peaks at $z\sim 0.35$.  {\it Top right:} The distribution of tangential separations between LRG pairs in Mpc$/h$. The corresponding angular distance at $z = 0.35$ in arcmin on the upper axis. {\it Bottom right}: The distribution of radial separations between LRG pairs. }
\label{f01}
\end{center}
\end{figure*}

We construct the LRG pair catalog from this central LRG catalog by finding all neighbouring LRG's within a tangential distance of 6--10 $h^{-1} \mathrm{Mpc}$ and within a proper radial distance of $\pm$ 6 $h^{-1} \mathrm{Mpc}$.  The resulting catalog has 262,864 LRG pairs to redshifts $z \sim 0.4$.  Their redshift and separation distributions are shown in Figure~\ref{f01}.

\subsection{Planck $y$ map}
The \planck\ tSZ map is one of the datasets provided in the \planck\ 2015 data release.  It is available in HEALpix\footnote{http://healpix.sourceforge.net/} format with a pixel resolution of $N_{\rm side}$ = 2048.  Two types of $y$ map are publicly available: MILCA and NILC, both of which are based on multi-band combinations of the \planck\ band maps \citep{Planck2016-XXII}.  Our analysis is based on the MILCA map, but we obtain consistent results with the NILC map.

The 2015 data release also provides sky masks suitable for analyzing the $y$ maps, including a point source mask and galactic masks that exclude 40, 50, 60 and 70\% of the sky.  We combine the point source mask with the 40\% galactic mask which excludes $\sim$50\% of the sky.   The mask is applied during the stacking process: for a given LRG pair, masked pixels in the $y$ map near that pair are not accumulated in the stacked image.

\subsection{Simulations\label{bahamas}}
To compare our results with theory, we analyze the BAHAMAS suite of cosmological smoothed particle hydrodynamics (SPH) simulations \citep{McCarthy2017} in the same manner as the data.  The BAHAMAS suite is a direct descendant of the OWLS \citep{Schaye2010} and cosmo-OWLS projects \citep{Brun2014, Daalen2014, McCarthy2014}.  The simulations reproduce a variety of observed gas features in groups and clusters of galaxies in the optical and X-ray bands.  The BAHAMAS suite consists of periodic box hydrodynamical simulations,  the largest of which have volumes of (400 $h^{-1} \mathrm{Mpc}$)$^3$ and contain $1024^3$ each of baryonic and dark matter particles.   The suite employs two different cosmological models: the \wmap9 cosmology \citep{Hinshaw2013} with
\begin{align*}
& \lbrace \Omega_{\rm m}, \Omega_{\rm b}, \Omega_{\Lambda}, \sigma_8 , n_{\rm s} , h \rbrace = \\ 
& \lbrace 0.2793, 0.0463, 0.7207, 0.821, 0.972, 0.700 \rbrace,
\end{align*}
and the \planck\ 2013 cosmology \citep{Planck2014-I} with
\begin{align*}
& \lbrace \Omega_{\rm m}, \Omega_{\rm b}, \Omega_{\Lambda}, \sigma_8 , n_{\rm s} , h \rbrace = \\
& \lbrace 0.3175, 0.0490, 0.6825, 0.834, 0.9624, 0.6711 \rbrace.
\end{align*}
We have four realizations based on the \wmap9 cosmology and one based on the \planck\ 2013 cosmology.

From each realization, 10 nearly-independent mock galaxy catalogs are generated on 10 nearly-independent light cones, and 10 corresponding $y$ maps are generated from the hot gas component of the simulation \citep{McCarthy2014}. Each of the light cones contain about one million galaxies out to $z \sim 1$, and each spans a $10^{\circ} \times 10^{\circ}$ patch of sky.  To compare with data, we convolve the simulated $y$ maps with a Gaussian kernel of 10 arcmin, FWHM, corresponding to the \planck\ beam.

%%%%%%%%%%%%%%%%%%%%%%%%%%%%%%%%%%%%%%%%%%%%%%%%%%%%%%%%%%%%%
%
%                   Pair analysis
%
%%%%%%%%%%%%%%%%%%%%%%%%%%%%%%%%%%%%%%%%%%%%%%%%%%%%%%%%%%%%%

\section{Data Analysis}
In this section we describe our procedure for stacking a $y$ map against LRG pairs, we estimate and subtract the signal associated with single galaxy haloes, and we estimate the uncertainty of our final result.

\subsection{Stacking on LRG pairs}
\label{sec:stacking}
The angular separation between LRG pairs in our catalog ranges between 27 and 203 arcmin.  For each pair in the catalog, we follow \cite{Clampitt2016} and form a normalized 2-dimensional image coordinate system, ($X,Y$), with one LRG placed at ($-1,0$) and the other placed at ($+1,0$).   The corresponding transformation from sky coordinates to image coordinates is also applied to the $y$ map and the average is taken over all members in the catalog.  The mean tSZ signal in the annular region $9 < r < 10$ ($r^2 = X^2+Y^2$) is subtracted as an estimate of the local background signal.

\begin{figure}
\includegraphics[width=\linewidth]{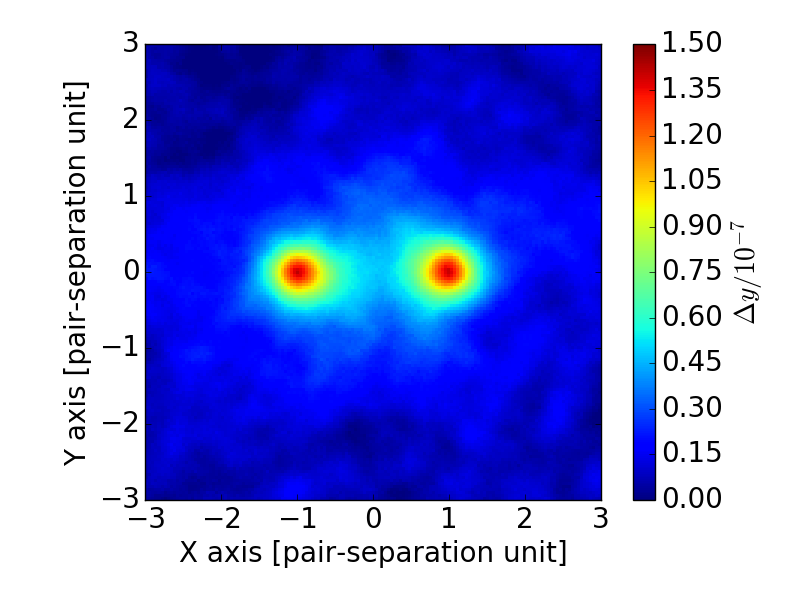}
\includegraphics[width=0.9\linewidth]{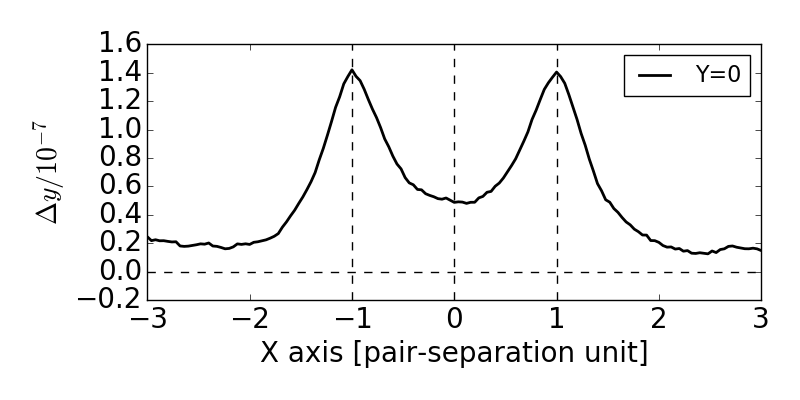}
\caption{{\it Top}: The average \planck\ $y$ map stacked against 262,864 LRG pairs in a coordinate system where one LRG is located at $(X,Y)=(-1,0)$ and the other is at $(X,Y)=(+1,0)$.  The square region, $-3 < X < +3$ and $-3 < Y < +3$, comprises 151 $\times$ 151 pixels.  {\it Bottom}: The corresponding $y$ signal along the X axis.}
\label{f02}
\end{figure}

The top panel of Figure~\ref{f02} shows the average $y$ map stacked against 262,864 LRG pairs over the domain $-3 < X < +3$ and $-3 < Y < +3$, and the lower panel shows a slice of this map at $Y=0$.  Not surprisingly, the average signal is dominated by the halo gas associated with the individual LRGs in each pair.  The peak amplitude of this signal is $\Delta y \sim 1.2 \times 10^{-7}$.

\subsection{Subtracting the halo contribution}
\label{sec:halo}
\begin{figure}
\includegraphics[width=\linewidth]{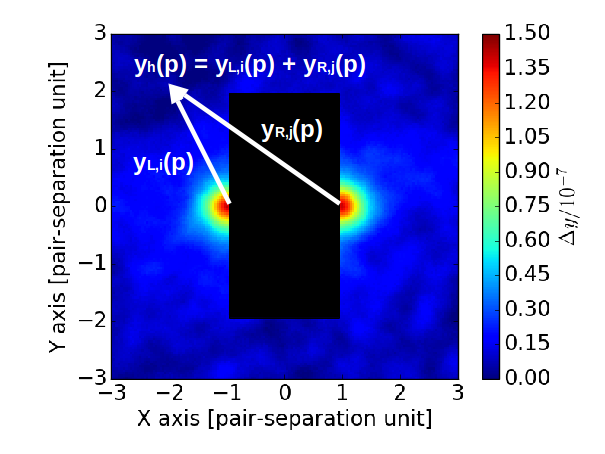}
\caption{Illustration to estimate the radial profile of each single halo. The $y$ value at a pixel $p$, $y_h(p)$, is the sum of $ith$ radial bin centred on the left LRG at ($-1,0$), $y_{L,i}(p)$ , and $jth$ radial bin centred on the right LRG, $y_{R,j}(p)$ at ($+1,0$). To avoid biasing the profiles with non-circular filament signal, the central region $-1 < X < +1$ and $-2 < Y < +2$ is excluded. }
\label{f03}
\end{figure}

\begin{figure}
\includegraphics[width=\linewidth]{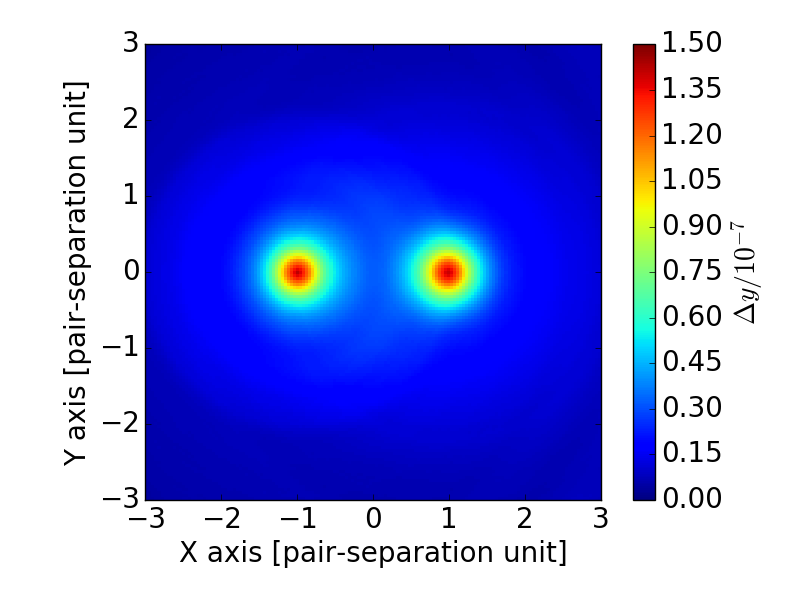}
\includegraphics[width=0.9\linewidth]{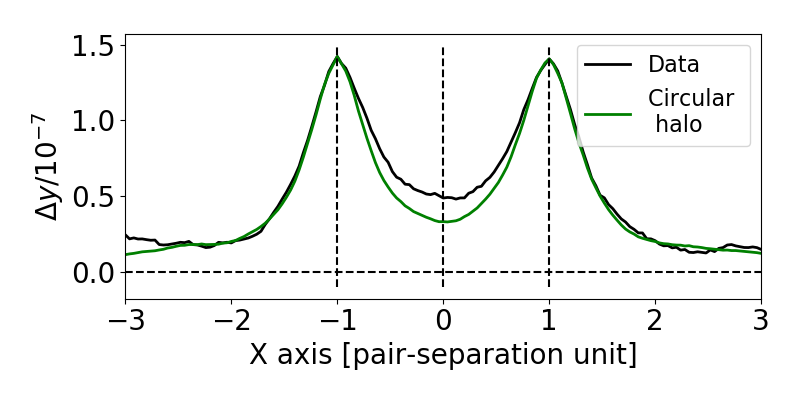}
\caption{{\it Top}: The circular halo profiles estimated to the map in Figure~\ref{f02}.  {\it Bottom}: The radial profile of the left and right halos shown above.}
\label{f04}
\end{figure}

We estimate the average contribution from single LRG halos as follows.  Since we have selected central LRGs for our pair catalog, we assume that the average single-halo contribution is circularly symmetric about each LRG in a pair.   To determine the radial profile of each single halo, we analytially solve for the map (indexed by pixel $p$) to a model of the form 
\beq
y_h(p)  = y_{L,i}(p) + y_{R,j}(p),
\eeq
where $y_{L,i}$ is the single-halo signal in the $ith$ radial bin centred on the ``left'' LRG at ($-1,0$), $y_{R,j}$ is the single-halo signal in the $jth$ radial bin centred on the ``right'' LRG at ($+1,0$), 
and $p=p(X,Y)$ is a pixel on the map as shown in Figure~\ref{f03}. When solving it, we choose radial bins of size $\Delta r = 0.02$, and we weight the map pixels uniformly.  To avoid biasing the profiles with non-circular filament signal, we exclude the central region $-1 < X < +1$ and $-2 < Y < +2$.  Figure~\ref{f04} shows the resulting profiles for the LRG halos.

\begin{figure}
\includegraphics[width=\linewidth]{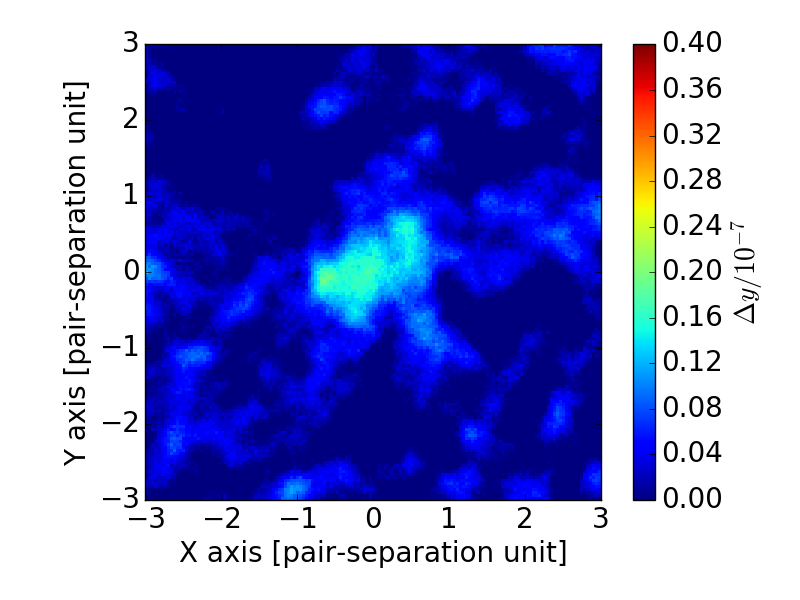}
\includegraphics[width=0.9\linewidth]{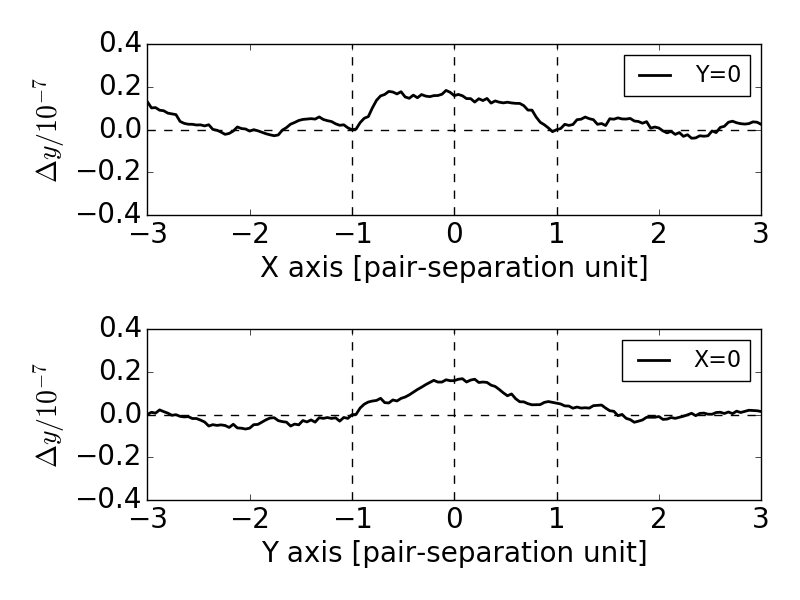}
\caption{{\it Top}: The residual $y$-map after the radial halo signals are subtracted.  {\it Bottom}: The residual tSZ signal along the $X$ and $Y$ axes.}
\label{f05}
\end{figure}

Figure~\ref{f05} shows the residual $y$ map after subtracting the circular halo profiles shown in Figure~\ref{f04} (note the change in colour scale from Figures~\ref{f04} and \ref{f05}).  The bright halo signals appear to be cleanly subtracted, while a residual signal between the LRGs is clearly visible.  The lower panels of Figure~\ref{f05} show the residual signal in horizontal ($Y=0$) and vertical ($X=0$) slices through the map.  The shape of the signal is consistent with an elongated filamentary structure connecting average pairs of central LRGs.  The mean residual signal in the central region, $-0.8 < X < +0.8$ and $-0.2 < Y < +0.2$, is found to be $\Delta y = 1.31 \times 10^{-8}$.

\subsection{Null tests and error estimates}
\begin{figure}
\includegraphics[width=\linewidth]{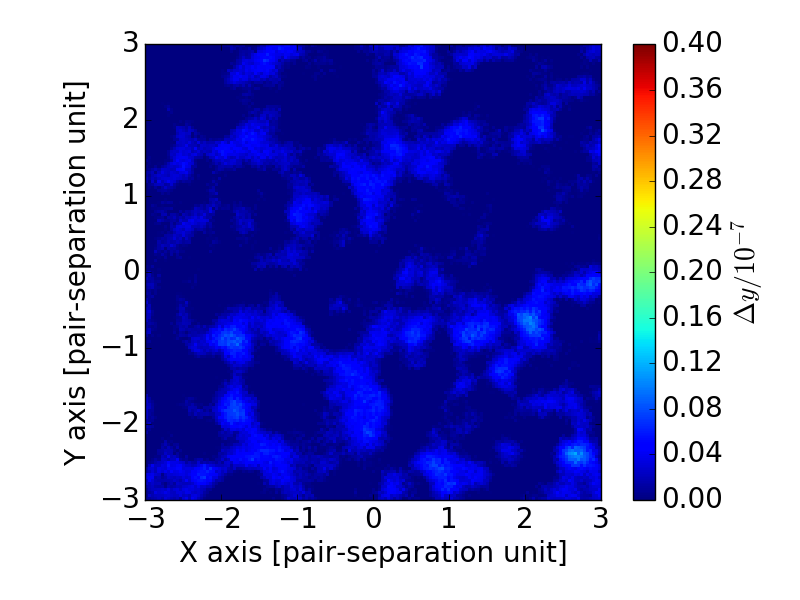}
\includegraphics[width=0.9\linewidth]{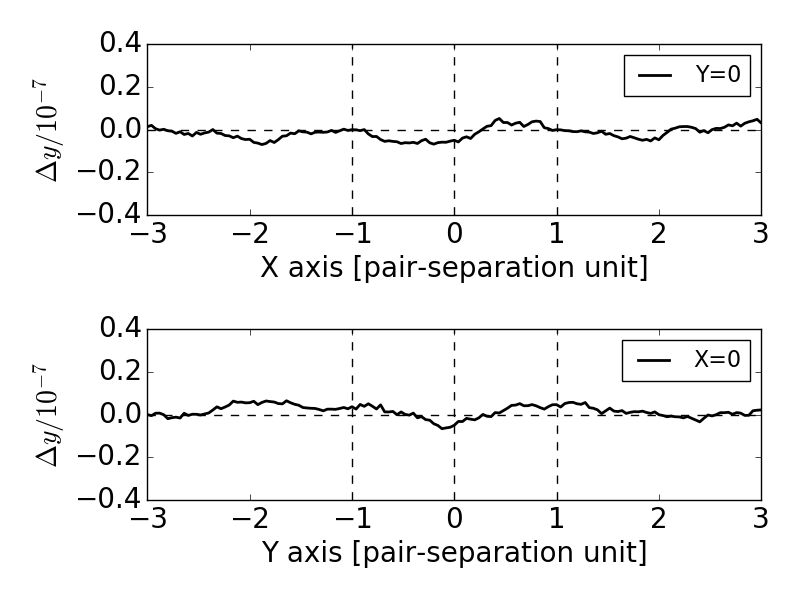}
\caption{{\it Top}: A sample null map obtained by stacking the $y$ map against the LRG pairs that were rotated in galactic longitude by random amounts.  {\it Bottom}: The tSZ signal along the X and Y axes of the $y$ map shown above.}
\label{f06}
\end{figure}

To assess the reality of the residual signal and estimate its uncertainty, we perform two types of Monte Carlo-based null tests.  In the first test, we rotate the centre of each LRG pair by a random angle in galactic longitude, for example, the centre of one LRG pair is rotated from $[l,b] = [10^{\circ}, 60^{\circ}]$ to $[l,b] = [150^{\circ}, 60^{\circ}]$.  (Keeping the latitude fixed keeps the galactic foreground level in the $y$ map approximately fixed.)  We then stack the $y$ map against the set of rotated LRG pairs, and we repeat this stacking of the full catalog 1000 times to determine the $rms$ fluctuations in the background (and foreground) sky.  Figure~\ref{f06} shows one of the 1000 rotated, stacked $y$ maps: the map has no discernible structure.  We can use this ensemble of maps to estimate the uncertainty of the filament signal quoted above.  Taking the same region used before ($-0.8 < X < +0.8$ and $-0.2 < Y < +0.2$), we find that the ensemble of null maps has a mean and standard deviation of $\Delta y = (-0.03 \pm 0.24) \times 10^{-8}$ in Figure~\ref{f08}.  Since the average signal in this null test is consistent with zero, we cautiously infer that our estimator is unbiased, however, we present another test in the following. 

The second null test is to stack the $y$ map against ``pseudo-pairs'' of LRGs: that is, pairs of objects that satisfy the transverse separation criterion, but which have a large separation along the radial direction.  Such pairs are not expected to be connected by filamentary gas.  We generate a pseudo-pair catalog as follows: for each pair in the original catalog, we pick one of the two members at random, then pick a new partner LRG that is located within 6--10 $h^{-1} \mathrm{Mpc}$ of it in the transverse direction, but which is {\em more} than 30 $h^{-1} \mathrm{Mpc}$ away in the radial direction.  We select the same number of pairs meeting this criterion as in default LRG pair catalog, so that the stacked image is of approximately the same depth. We repeat this selection 1000 times to generate an ensemble of pseudo-pair catalogs.

\begin{figure}
\includegraphics[width=\linewidth]{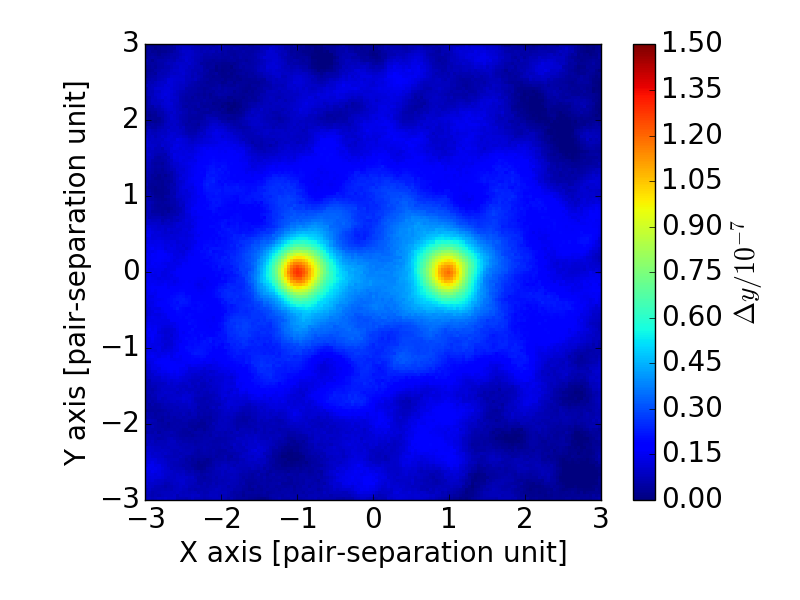}
\includegraphics[width=\linewidth]{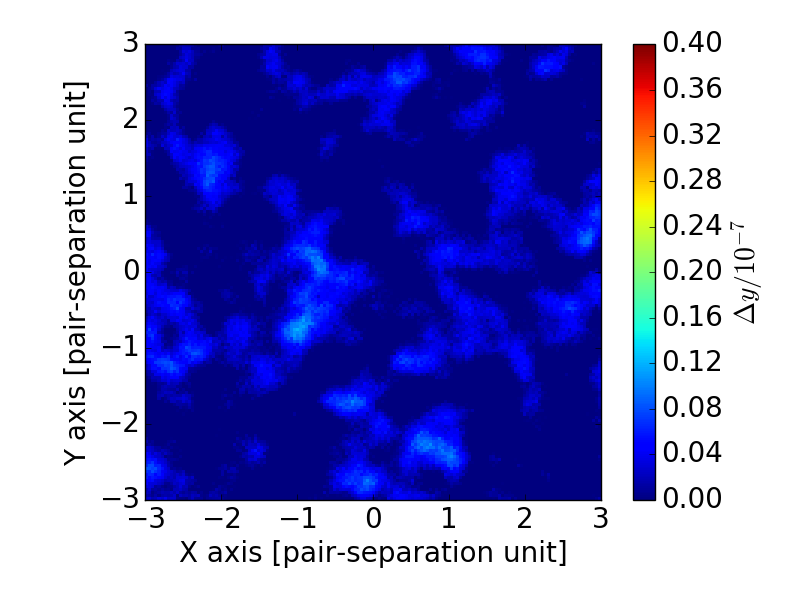}
\includegraphics[width=0.9\linewidth]{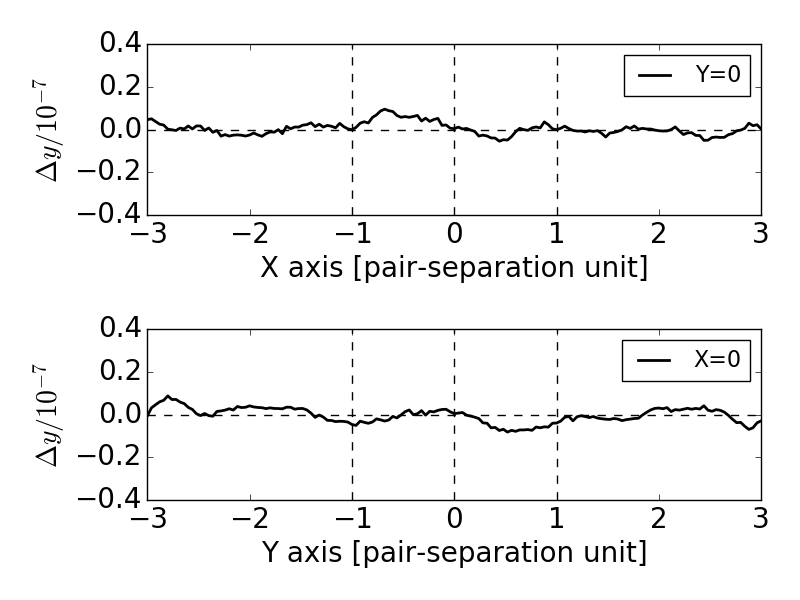}
\caption{{\it Top}: An average $y$ map stacked against a catalog of LRG pseudo pairs (see text for a definition).  {\it Middle}: The residual $y$ map after subtracting the circular halos from the above map, using the same procedure that was applied to the genuine pair stack.  {\it Bottom}: The tSZ signal along the X and Y axes of the residual map shown in the middle panel.}
\label{f07}
\end{figure}

\begin{figure}
\includegraphics[width=\linewidth]{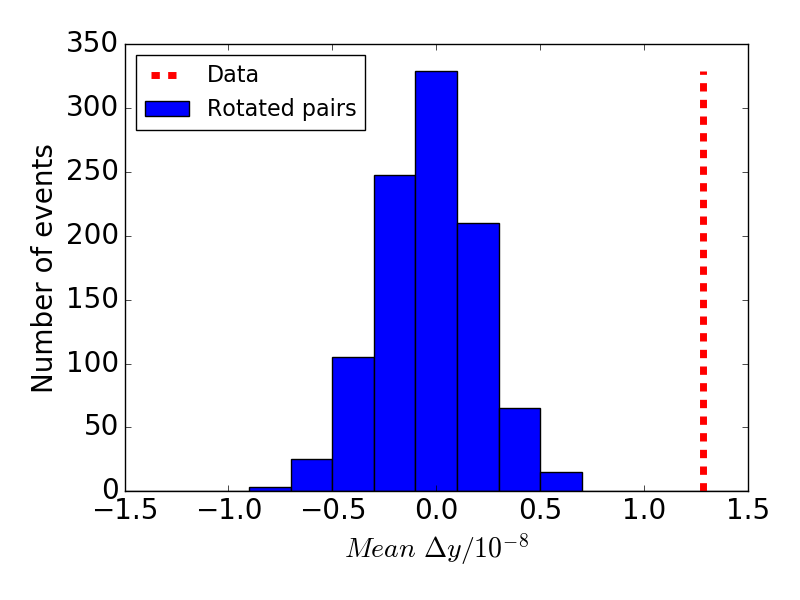}
\includegraphics[width=\linewidth]{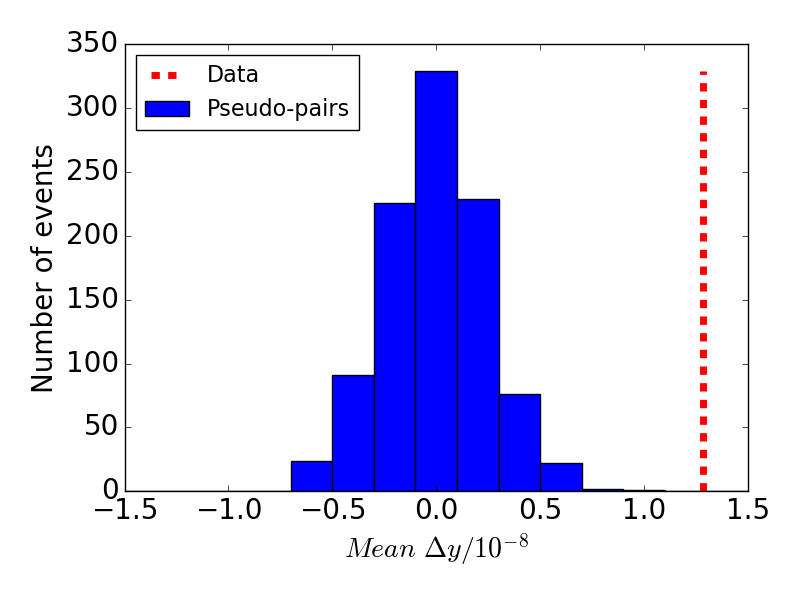}
\caption{{\it Top}: The mean filament amplitude measured in 1000 randomly-rotated pair stacks.  {\it Bottom}: The same statistic evaluated in 1000 pseudo-pair stacks. Both tests are consistent with zero residual signal.  The mean filament amplitude measured in the data ($\Delta y = 1.31 \times 10^{-8}$) is indicated by red-dash line.}
\label{f08}
\end{figure}

The top panel of Figure~\ref{f07} shows an average $y$ maps stacked against one of the pseudo pair catalog realizations.  This map is similar to the genuine pair-stacked map shown in Figure~\ref{f02}, but with less apparent signal between the LRGs.  We perform the same single-halo model calculation described above and subtract it from the map with the result is shown in the middle panel of Figure~\ref{f07}.  As with the rotated null test above, this map shows no discernible structure.  To generate statistics, we repeat this test 1000 times: we find that the ensemble of null maps has a mean and standard deviation of $\Delta y = (0.00 \pm 0.25) \times 10^{-8}$ in Figure~\ref{f08}, virtually the same as with the rotated ensemble.  We adopt this standard deviation as the final uncertainty of the mean filament signal due to instrument noise, sky noise (i.e., cosmic variance and foreground rejection errors), and halo subtraction errors.

\section{Interpretation}
\label{sec-int}

\subsection{Systematic errors}
Some systematic effects that might enhance or diminish the tSZ signal in the filaments may exist. For example, in our analysis, although we assume that the average single-halo contribution is circularly symmetric about each LRG, one might speculate that the signal we detect between nearby LRG pairs is a result of tidal effects, in which single-object halos are elongated along the line joining the two objects.  However, considering that the $\sim$ 260,000 LRG pairs are made from $\sim$ 220,000 LRGs and each LRG has roughly 2 pairs (One is aligned between our LRG pairs and the other is not aligned),  the direction of the elongation is not obvious. While tidal effects must be present at some level, if they were the dominant explanation for the residual signal we see, we would expect the elongated halo structure to extend in {\em both} directions along the line joining the objects.  The fact that we see no significant excess signal outside the average pair suggests that tidal effects are not significant.

On the other hand, there are possible systematic effects that might reduce the tSZ signal in the filaments. For example, some of the LRG pairs are not connected by filaments, or some of the filaments may not be straight but curved. However, in a study of $N$-body simulations, \cite{Colberg2005} found that cluster pairs with separations $<5h^{-1}$ Mpc are always connected by dark matter filaments, mostly straight filaments.  Further, they found filaments connecting $\sim$85\% of pairs separated by 5--10 $h^{-1}$ Mpc and $\sim$70\% of pairs separated by 10--15 $h^{-1}$ Mpc. This effect would lower the average $y$-value in the stacked filament and/or broaden the shape at some level, but the simulation study implies that most LRG pairs should be connected by dark matter (and presumably gas) filaments, and that dilution is unlikely to be significant.

There may be contamination from asymmetric features around halos, due, for example, to filaments that do not join our LRG pairs. We estimate the magnitude of this effect assuming one additional filament randomly oriented around each halo.  The mean filament we observe between LRG pairs occupies roughly 10\% of the annular region around a halo, above a threshold of $\Delta y \sim 1.0 \times 10^{-8}$ ($0.2 < r < 1.8$, $r^2=(X-1)^2 + Y^2$ for the right LRG, and $r^2=(X+1)^2 + Y^2$ for the left LRG).  Thus, the circular halo profile in the annular region could be overestimated by $\Delta y \sim 1.0 \times 10^{-8}/10 = 1.0 \times 10^{-9}$, which is small compared to the mean filament signal. 

There may be also contamination from the CIB emission in the \planck\ $y$ map. The tSZ-CIB cross-correlation was carefully studied in \cite{planck2016-xxiii}. The Fig.14 in the \planck\ paper shows that the CIB contamination in the tSZ signal in the scale of 30 - 60 arcmin, which is the typical angular scale of filaments in our samples, amounts to less than 10 \% in the power spectrum.

The mean filament signal will be affected somewhat by beam dilution given the \planck\ angular resolution of 10$'$.  We estimate the magnitude of this effect using the BAHAMAS simulations.  With smoothed $y$ maps, the mean filament signal in the simulations is $\Delta y = 0.84 \times 10^{-8}$ (\S\ref{sec:comp-sim}) while in the un-smoothed maps it is $\Delta y = 1.00 \times 10^{-8}$.  In this model, beam smoothing dilutes the amplitude of the mean filament signal by $\sim$15\%.

\subsection{Unbound diffuse gas or bound gas in halos?}
\label{sec:unbound}
Is the signal we detect due to unbound diffuse gas outside of halos or to bound gas in halos between the LRG pairs?  We address this question in two ways: first, we compare the data to predictions based on the halo model, including the two-halo term to account for correlated halos; second, we construct and analyze a mock $y$ map using the halo model to ``paint'' $y$ signal at the location of each LRG in our catalog. 

\begin{figure}
\includegraphics[width=\linewidth]{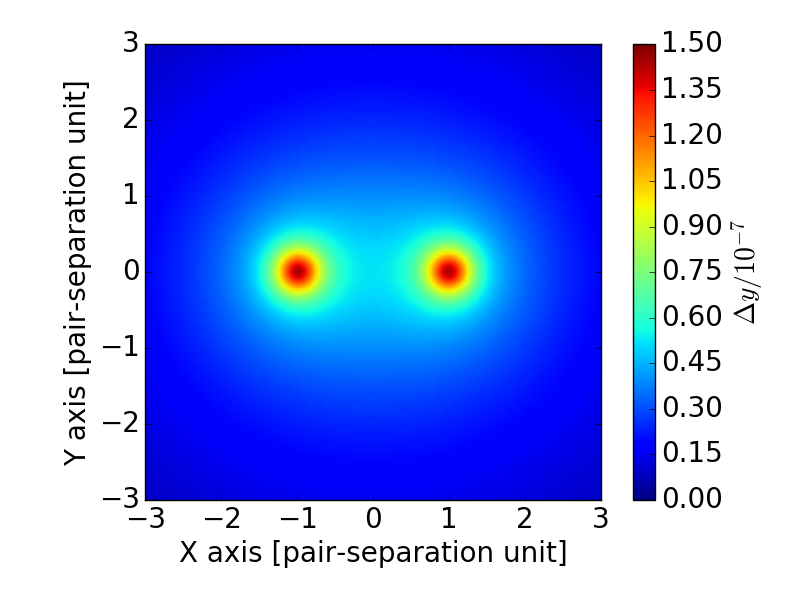}
\includegraphics[width=\linewidth]{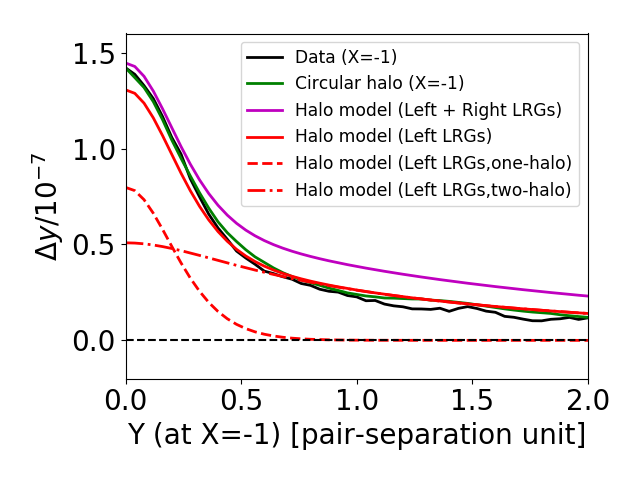}
\includegraphics[width=\linewidth]{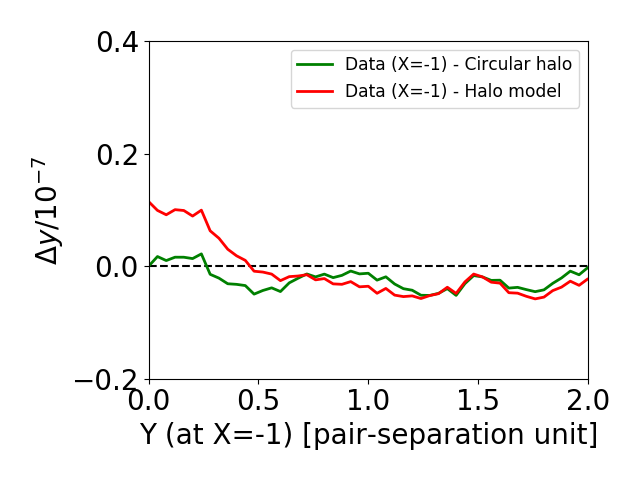}
\caption{{\it Top}: The model $y$ map constructed by the halo model prediction.  {\it Middle}: The halo model prediction of the $y$ signal along $X=-1$ compared to the data from Figure~\ref{f02} and our circular halo profile from Figure~\ref{f04}.  {\it Bottom}: The difference of the $y$ signal along $X=-1$ between the data and our circular halo profile and also halo model prediction.}
\label{f09}
\end{figure}

The results of the first test are shown in Figure~\ref{f09}.  The top panel shows the model map constructed using the halo model and the Universal Pressure Profile (UPP), which has an analytical formulation given by \cite{Nagai2007} for the generalised Navarro-Frenk-White (GNFW) profile \citep{Navarro1997}, to calculate the expected signal from individual halos. For the parameters in the GNFW, we adopt the best-fit values of $[P_{0},c_{500}, \gamma, \alpha, \beta] = [6.41,1.81,0.31,1.33,4.13]$, estimated using \planck\ tSZ and XMM-Newton X-ray data in \cite{Planck2013IR-V}. The method we use to construct the halo model profiles requires careful calibration of the stellar mass / halo mass relation, as detailed in \cite{tanimura2017phdt}.   In that paper, we showed that the average measured profile of $y$ in single LRGs agrees well with halo model predictions {\em as long as one accounts for the two-halo contribution due to correlated systems along the line of sight}.  The middle panel of Figure~\ref{f09} shows a number of different profile slices to aid interpretation.  To avoid confusion with the putative filament signal, we take vertical slices of the data and model at $X=-1$ for the left halo alone as follows: solid black shows the data $y$ profile in Figure~\ref{f02}; solid green shows the circular halo profile from Figure~\ref{f04}; solid red shows the prediction of the halo model using the same model parameters we used to successfully match the single LRG profiles \citep{tanimura2017phdt}.   (The red dashed curves show how the halo model breaks down into one-halo and two-halo contributions.) The bottom panel of Figure~\ref{f09} shows the difference of the $y$ signal along $X=-1$ between the data and our circular halo profile and also halo model prediction. All three of these profiles are in good agreement, demonstrating that the halo model is successfully able to predict the halo signal in directions away from the filament.  But it is important to note that the red curve includes a two-halo contribution: without it, the model would fall short of the data in the outer regions of the halo.

The purple curve in Figure~\ref{f09} shows the prediction of the halo model for both left and right LRGs combined.  The over-prediction of this model arises because the two-halo term falls off slowly on scales comparable to the separation between the LRGs, so it is effectively being counted twice when we form the pair-stack map from the two profiles.  To correctly apply the halo model in this case, we would need to develop the formalism of a three-halo contribution, which is beyond the scope of this paper.

\begin{figure}
\includegraphics[width=0.50\linewidth]{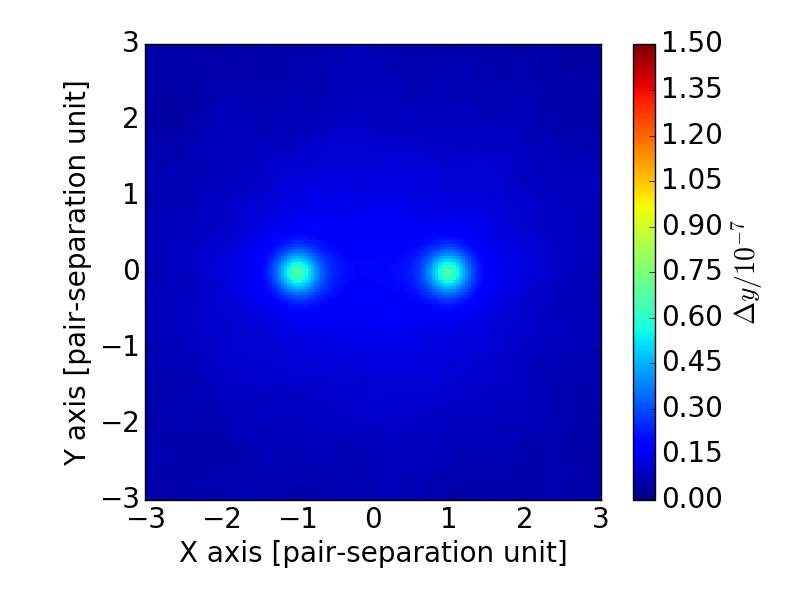}
\includegraphics[width=0.50\linewidth]{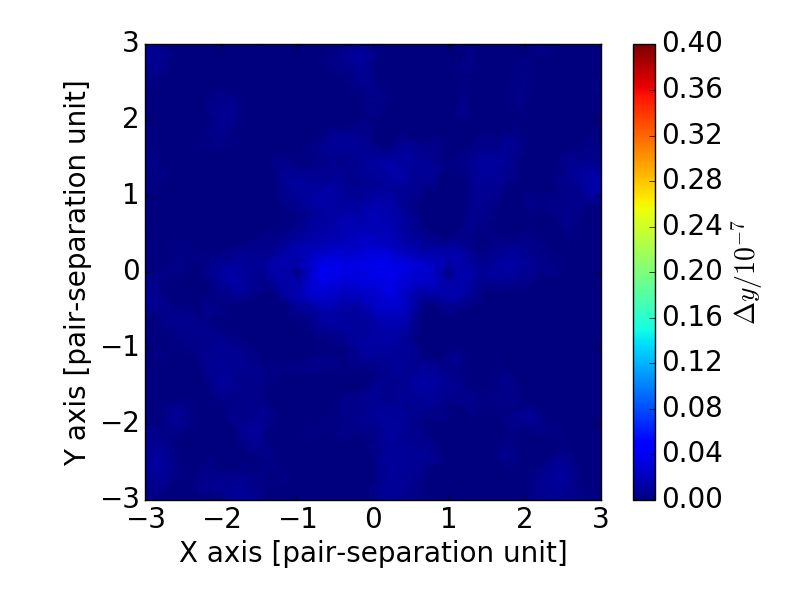}
\caption{{\it Left}: The single-halo model $y$ map described in the text is stacked against the same 262,864 LRG pairs as in the data analysis. {\it Right}: The residual model $y$ map after subtracting the circular halos from the map at {\it left}, using the same procedure that was applied to the genuine pair stack.}
\label{f10}
\end{figure}

In the second test, we simulate a model $y$ map using only bound gas in the SDSS DR12 LRGs with $10^{10} M_{\odot} < M_{\ast} < 10^{12} M_{\odot}$ and $0. < z < 0.8$ and compare the result with \planck\ $y$ map. To make the single-halo model $y$ map, we select ``central`` LRGs described in \S\ref{sec:lrg}, which leaves $\sim$1,100,000 LRGs, and estimate the halo masses of the LRGs with the stellar-to-halo mass relation of \cite{Coupon2015}, estimated in the CFHTLenS/VIPERS field by combining deep observations from the near-UV to the near-IR, supplemented by $\sim$70,000 secure spectroscopic redshifts, and analyzing galaxy clustering, galaxy-galaxy lensing and the stellar mass function. Then we locate $y$ profiles within the virial radius ($r < r_{200}$) of the LRG halos on the map using the UPP. We convolve the model $y$ map with a 10 arcmin FWHM Gaussian kernel to match the \planck\ map resolution. We perform the same analysis on the model $y$ map other than the actual \planck\ map. The peak $y$ valus of the LRG halos in Figure~\ref{f10} is dimmer than the $y$ map in Figure~\ref{f02} since we only include the contribution within the virial radius of the LRG halos and in addition, no sub-halos are included. After the circular halo subtraction, we obtain $\Delta y = 0.29 \times 10^{-8}$ from the bridge region. Moreover, we simulate a model $y$ map including $y$ profiles within $r < 3 \times r_{200}$ of the LRG halos and the result is $\Delta y = 0.36 \times 10^{-8}$. These results suggest that most of the $y$ signal we detect between the LRGs could originate in unbound diffuse gas, although the contributions from other types of galaxies should be present at some level. It would be also important to understand an extend of hot gas around halos, especially for low-mass halos, as \cite{mcgaugh2010} shows a depletion of baryons in low-mass halos relative to the cosmic fraction. 

\subsection{Gas properties}
\label{sec:gasprop}
We can estimate the physical conditions of the gas we detect by considering a simple, isothermal, cylindrical filament model of electron over-density with a density profile proportional to $r_c/r$, at redshift $z$.  The Compton $y$ parameter produced by the tSZ effect is given by
\beq
y = \frac{\sigma_{\rm T} k_{\rm B}}{m_{\rm e} c^2}  \int n_{\rm e} \, T_{\rm e} \, dl, 
\label{eq-y}
\eeq
where $\sigma_{\rm T}$ is the Thomson scattering cross section, $k_{\rm B}$ is the Boltzmann constant, $m_{\rm e}$ is the electron mass, $c$ is the speed of light, $n_{\rm e}$ is the electron number density, $T_{\rm e}$ is the electron temperature, and the integral is taken along the radial direction. In general, the electron density at position ${\bf x}$ may be expressed as 
\beq
n_{\rm e}({\bf x},z) = \overline{n}_{\rm e}(z)(1+\delta({\bf x})),
\eeq
where $\delta({\bf x})$ is the density contrast, and $\overline{n}_{\rm e}(z)$ is the mean electron density in the universe at redshift $z$,
\beq
\overline{n}_{\rm e}(z) = \frac{\rho_{\rm b} ({\it z})}{\mu_{\rm e} m_{\rm p}}, 
\eeq
where $\rho_{\rm b} (z) = \rho_{\rm c} \Omega_{\rm b} (1+z)^3$ is the baryon density at redshift $z$, $\rho_{\rm c}$ is the present value of critical density in the universe, $\Omega_{\rm b}$ is the baryon density in units of the critical density, $\mu_{\rm e} = \frac{2}{1+\chi} \simeq 1.14$ is the mean molecular weight per free electron for a cosmic hydrogen abundance of $\chi = 0.76$,  and $m_{\rm p}$ is the mass of the proton.

We can express the profile in the Comptonisation parameter as a geometrical projection of a density profile with $n_{\rm e}(r,z)$
\beq
y(r_{\bot}) = \frac{\sigma_{\rm T} k_{\rm B} T_{\rm e}}{m_{\rm e} c^2}  \int^{R}_{r_{\bot}} \frac{2r \, n_{\rm e}(r,z)}{\sqrt{r^2 - r_{\bot}^2}} \, dr, 
\eeq
where $r_{\bot}$ is the tangential distance from the filament axis on the map and $R$ is the cut-off radius of the filaments. Assuming negligible evolution of the filaments, and a constant over-density, $\delta_c$, at the core, out to $z = 0.4$ (the maximum redshift in our LRG sample),
\beq
n_{\rm e}(r=0,z) = \frac{n_{\rm e}(r=0,z) \, \overline{n}_{\rm e}(z)}{\overline{n}_{\rm e}(z)} = \delta_c \, \overline{n}_{\rm e}(0) \, (1+z)^3. 
\eeq
We consider three density profiles, 
\begin{eqnarray}
n_{\rm e}(r) &=& constant \,\, (r < r_c), \\
n_{\rm e}(r) &=& \frac{n_{\rm e}(0)}{\sqrt{1+(r/r_c)^2}} \,\, (r < 5r_c), \\
n_{\rm e}(r) &=& \frac{n_{\rm e}(0)}{1+(r/r_c)^2} \,\, (r < 5r_c), 
\end{eqnarray}
where $r_c$ is the core radius. To regularize the profiles, we adopt a cutoff radius of $5r_c$ for the second and third profile. Applying the profiles to the simulations, we find the best-fit density profile to follow $(r_c/r)$.

For this model, the predicted mean tSZ signal, $\overline{\Delta y}$, in the region $-0.8 < X < +0.8$, $-0.2 < Y < +0.2$, for the 262,864 filaments may be written as 
\beq
\overline{\Delta y} = 4.9 \times 10^{-8} \times \left(\frac{\delta_c}{10} \right) \left(\frac{T_{\rm e}}{10^7 \; \rm{K}} \right) \left(\frac{r_c}{0.5h^{-1} \; {\rm Mpc}} \right).
\eeq
Applying the observational constraint $\overline{\Delta y} = (1.31 \pm 0.25) \times 10^{-8}$, we have,
\beq
\delta_c \left(\frac{T_{\rm e}}{10^7 \; \rm{K}} \right) \left(\frac{r_c}{0.5h^{-1} \; {\rm Mpc}} \right) = 2.7 \pm 0.5.
\eeq
Assuming the same temperature and morphology estimates from the simulations studied in \S\ref{sec:comp-sim} below apply to the observational data, the mean filament over-density between LRG pairs is $\delta \sim 3.2 \pm 0.7$. This suggests that the gas in the filaments between LRGs has a relatively low density.

\section{Comparison with hydrodynamic simulations}
\label{sec:comp-sim}
We compare our results to simulations.  To do so, we analyze light cones from the BAHAMAS suite of simulations (\S\ref{bahamas}) as we did the real data.  For each cosmology, we construct simulated LRG pairs by selecting central galaxies with the same separation criteria as the real data.  We invoke a stellar mass threshold such that the mean stellar mass of the sample matches the mean of the data.  The resulting catalog has 242,669 pairs.  Prior to stacking, we also convolve the simulated $y$ map in each light cone with a 10 arcmin FWHM Gaussian kernel to match the \planck\ map resolution.  After stacking and radial halo subtraction, we find the residual tSZ signal between central galaxy pairs to be $\Delta y = (0.84 \pm 0.24) \times 10^{-8}$ with the \wmap 9 cosmology.  The uncertainty is estimated by drawing 1000 bootstraps samples from among the 40 light cones.  We have also analyzed the simulations based on the \planck\ 2013 cosmology and find $\Delta y = (1.14 \pm 0.33) \times 10^{-8}$.  However, this model only has one realization of the initial conditions, instead of four, so it has a larger uncertainty than the \wmap 9 estimate. In conclusion, we find a slightly lower, but marginally consistent result from the simulations for both cosmologies.

The comparison of the simulations to the data is not entirely straightforward because of possible selection effects. In particular, the methods for estimating stellar mass in these two systems are different.  The data estimates we use are based on the principal component method in \cite{Chen2012}, which are, on average, $\sim$0.2 dex higher than those based on spectro-photometric model fitting \citep{Maraston2013}.  The simulation estimates we use are based on directly counting the baryonic mass within 30 kpc of a given central galaxy.  As noted in \S\ref{sec:lrg}, we adopt a stellar mass threshold of $10^{11.3}$ $M_{\sun}$ for the data.  In order to match the {\em mean} stellar mass of the simulation population, we must adopt a stellar mass threshold of $10^{11.2}$ $M_{\sun}$.  This procedure produces the same peak $y$ values at the center of each mean halo: data and simulation.  We believe this selection should produce comparable filament amplitudes.

In addition, we examine four independent realizations of the \wmap 9 cosmology and find that the mean residual tSZ signal between central galaxy pairs varies by a factor of $\sim$5, from $\Delta y = 0.26 \times 10^{-8}$ to $1.37 \times 10^{-8}$. This reflects the fact that cosmic variance is larger in the simulations than in the data due to their limited volume.

\begin{figure*}
\begin{center}
\begin{minipage}{0.4\linewidth}
\includegraphics[width=\linewidth]{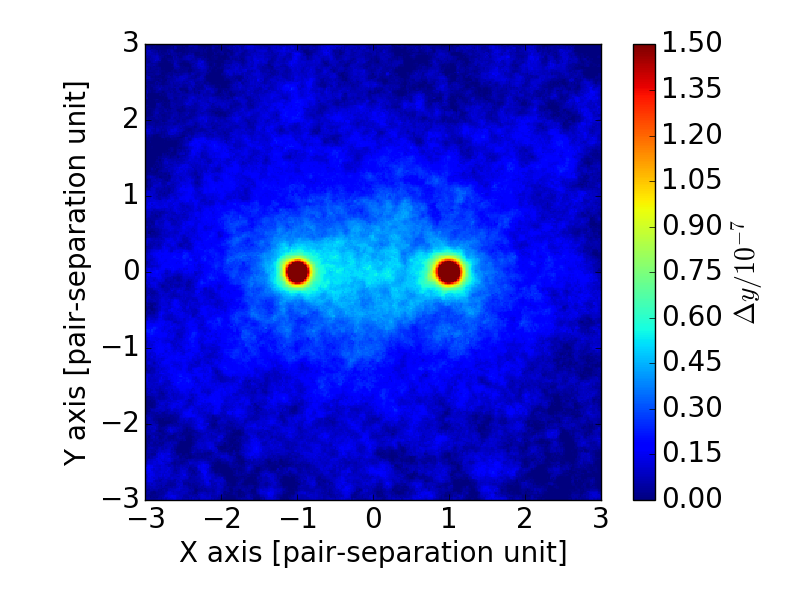}
\end{minipage}
\begin{minipage}{0.4\linewidth}
\includegraphics[width=\linewidth]{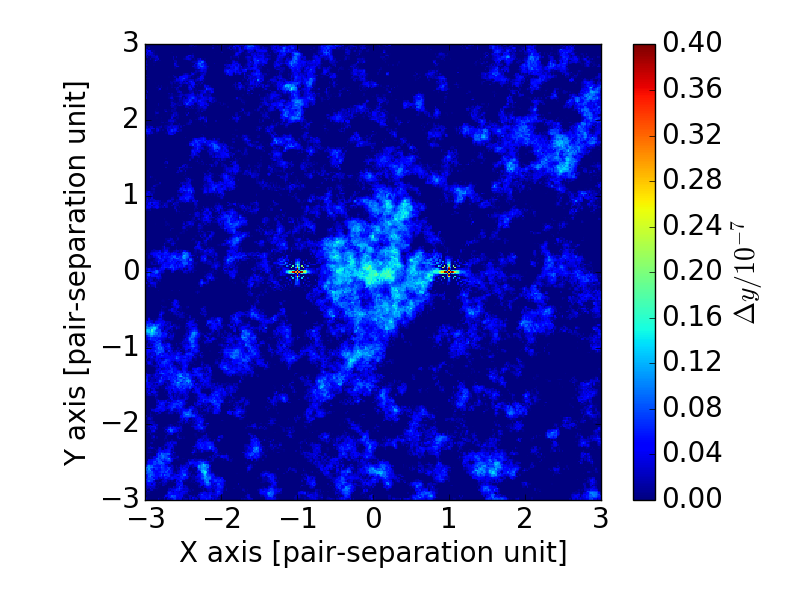}
\end{minipage}
\begin{minipage}{0.4\linewidth}
\includegraphics[width=\linewidth]{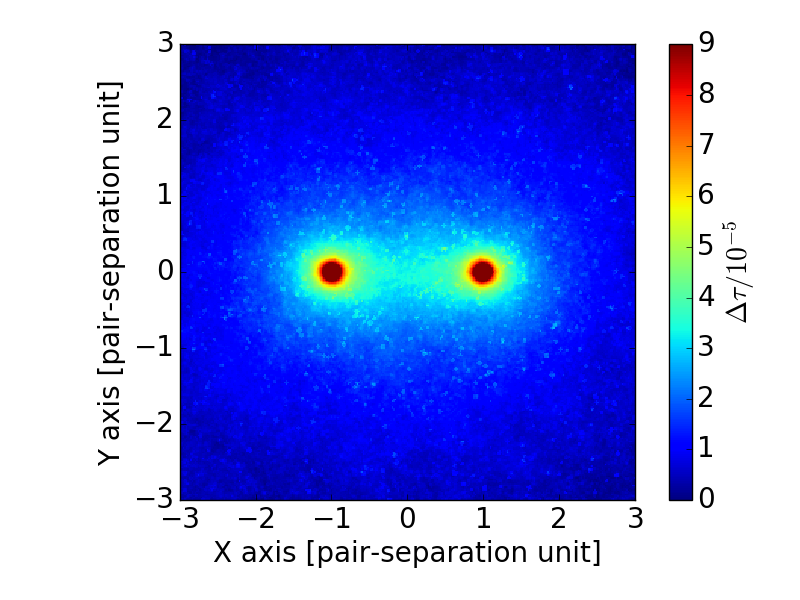}
\end{minipage}
\begin{minipage}{0.4\linewidth}
\includegraphics[width=\linewidth]{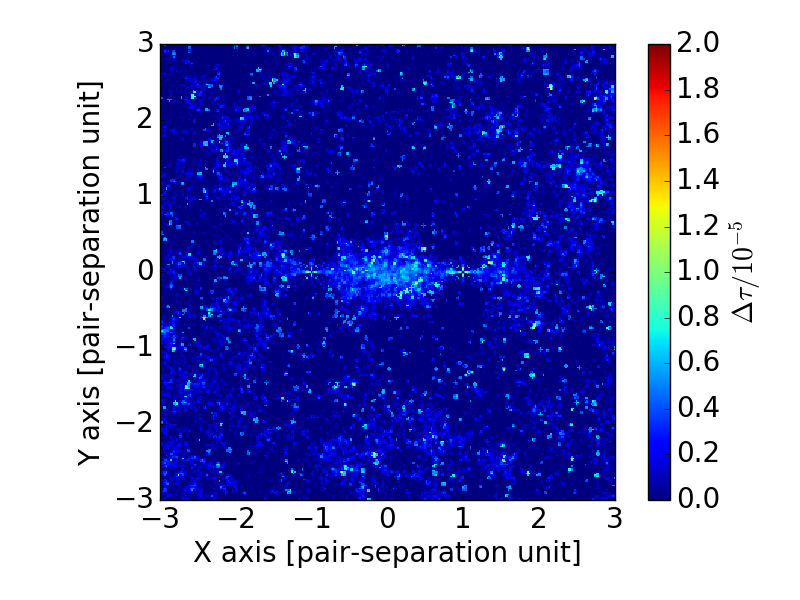}
\end{minipage}
\begin{minipage}{0.4\linewidth}
\includegraphics[width=\linewidth]{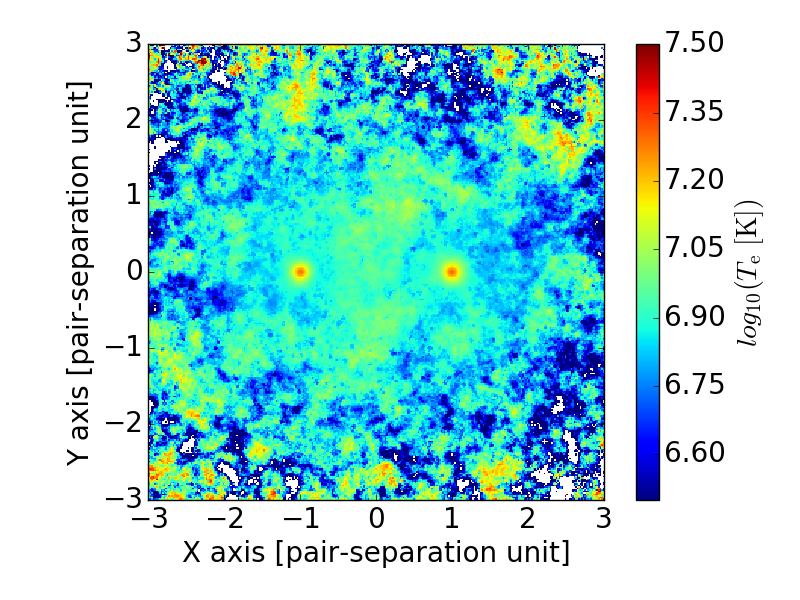}
\end{minipage}
\caption{{\it Top left}: The stacked $y$ map of the central galaxy pairs from the BAHAMAS simulations, at 10 arcsecond angular resolution.  {\it Top right}: The same $y$ map after the circular halos are subtracted.  {\it Middle left}: The stacked $\tau$ map for the same pair sample as above.  {\it Middle right}: The same $\tau$ map after circular halo subtraction.  {\it Bottom}: The electron-density-weighted electron temperature map, $T_{\rm e}$, on a $\log_{10}$ scale.}
\label{f11}
\end{center}
\end{figure*}

With the caveats noted above, we can use the simulations to further probe the physical conditions in the observed filaments using the high-resolution simulated maps.  In Figure~\ref{f11}, we separately examine the electron over-density and temperature in the stacked simulation data.  For each light cone in our simulation box, we form optical depth and electron temperature maps,
\beq
\Delta \tau = \tau - \overline{\tau} \, \left( with \ \tau = \sigma_{\rm T} \int n_{\rm e} \, dl \right),
\label{eq-tau}
\eeq
\beq
T_{\rm e} =  \frac{\Delta y}{\Delta \tau \times \frac{k_{\rm B}}{m_{\rm e} c^2}},
\label{eq-Te}
\eeq
where $\overline{\tau}$ is the mean value at the local background described in \S\ref{sec:stacking}.  We fit a variety of density profile models (\S\ref{sec:gasprop}) to the stacked $\tau$ map, and find the best-fit profile to follow $(r_c/r)$, where $r$ is the perpendicular distance from the cylinder axis, and $r_c = 0.5 h^{-1} \mathrm{Mpc}$, where $r_c$ is the core radius of the density profile.  Assuming this profile, the best-fit central over-density is $\delta = 1.5 \pm 0.4$ for \wmap 9 cosmology. From the $T_{\rm e}$ map, the mean electron-density-weighted temperature of the electron gas in the filament region is found to be $(0.82 \pm 0.06) \times 10^7$ K. 

\section{Discussion}
Other groups have studied filamentary gas in the large scale structure.  We compare and contrast those results to ours as follows.

The \planck\ Team \citep{Planck2013IR-VIII} studied the gas between the merging Abell clusters A399 and A401, which have a tangential separation of 3 $h^{-1} \mathrm{Mpc}$.  Using a joint analysis of \planck\ tSZ data and \rosat\ X-ray data, they estimate a gas temperature of $kT = 7.1 \pm 0.9$ keV ($T \sim 8.2 \times 10^7$ K), and a central electron density of $n_{\rm e} = (3.72 \pm 0.17) \times 10^{-4}$ cm$^{-3}$ ($\delta \sim 1500$). \cite{Bonjean2018} obtains a similar result of $n_{\rm 0} = (4.3 \pm 0.7) \times 10^{-4}$ cm$^{-3}$ for the gas density in the filament between A399 and A401 using the \planck\ tSZ data and \suzaku\ X-ray data \citep{Fujita2008}. Assuming a filament diameter of 1.0 Mpc with a cylindrical shape, it corresponds to $y \sim 10^{-5}$.  This high density and temperature may be because the filaments in merger systems have been shock-heated and compressed more than normal. 

Using XMM-Newton observations, \cite{Werner2008} study the gas properties in a filament connecting the massive Abell clusters A222 and A223, at redshift $z \sim 0.21$.   Assuming a separation of 15 Mpc, they find $kT = 0.91 \pm 0.25$ keV ($T \sim 1.1 \times 10^7$ K) and $n_{\rm e} = (3.4 \pm 1.3) \times 10^{-5}$ cm$^{-3}$ ($\delta \sim 150$). In addition, \cite{Eckert2015} find filamentary structures around the galaxy cluster Abell 2744, at $z \sim 0.3$, and estimate a gas temperature of $T \sim 10^7$ K, and an over-density of $\delta \sim 200$ on scales of 8 Mpc.  Assuming a filament diameter of 1.0 Mpc with a cylindrical shape, it corresponds to $y \sim 10^{-7}$, which is one order of magnitude higher than our result. 

In a study with targets in our mass range, but using the CFHTLenS mass map, \cite{Epps2017} examined the weak lensing signal between pairs of SDSS-III/BOSS LRG's.  They find a mass of $(1.6 \pm 0.3) \times 10^{13} M_{\odot}$ for a stacked filament region of 7.1 $h^{-1} \mathrm{Mpc}$ long and 2.5 $h^{-1} \mathrm{Mpc}$ wide. Assuming a uniform-density cylinder, they estimate $\delta \sim 4$ in the filaments, consistent with our result.

A simulation study, \cite{Colberg2005} examined 228 filaments between pairs of galaxy clusters in a $\Lambda$CDM $N$-body simulation produced by \cite{Kauffmann1999}. They identify straight mass filaments longer than 5 $h^{-1}$ Mpc, normalize the length of each filament to unity, and find the average density of matter contained within 2 $h^{-1}$ Mpc of the (normalized) filament axis to be $\delta \sim 7$.  This is somewhat higher than our estimate of $\delta = 3.2 \pm 0.7$, however, the following factors may compromise this comparison.  

1) They select halos with masses larger than $10^{14} M_{\odot}$, whereas we select LRGs with the stellar masses larger than $10^{11.3} M_{\odot}$. According to the stellar mass / halo mass (SHM) relation used in \cite{Planck2013IR-XI}, this corresponds to halo masses with $M_{200} \sim (5-7) \times 10^{13} M_{\odot}$, and may include lower mass systems, given the scatter in the SHM relation.  \cite{West1995} suggest that the masses of filaments are correlated with the masses of the halos associated with them. Thus, this selection of low-mass halo pairs can result in lower-mass filaments, and hence a lower tSZ signal.

2) The $\Lambda$CDM simulation tracks dark matter, whereas we analyze hydrodynamic simulations which include baryonic effects such as radiative cooling, star formation, SN feedback and AGN feedback. The baryonic gas in the filaments may not trace the dark matter faithfully. 

3) They examine filaments between cluster pairs separated by 5 $h^{-1}$ to 25 $h^{-1}$ Mpc, whereas we study smaller separations of 6--10 $h^{-1}$ Mpc.

\section{Conclusion}
Using the \planck\ Sunyaev-Zel'dovich (tSZ) map and the SDSS DR12 catalog of Luminous Red Galaxies (LRGs), we search for warm/hot gas filamentary gas between pairs of LRGs by stacking the $y$ map on a grid aligned with the pairs.  We detect a strong signal associated with the LRG host halos and subtract that using a circularly symmetric model. We detect a statistically significant residual signal and draw the following conclusions.

\begin{itemize}
\item The residual tSZ signal in the region between LRG pairs is $\Delta y = (1.31 \pm 0.25) \times 10^{-8}$, with a 5.3$\sigma$ significance.  Assuming a simple, isothermal, cylindrical filament model of electron over-density with a radial density profile proportional to $r_c/r$ (as determined from simulations), we constrain the physical parameters of the gas in the filaments to be 
\beq
\delta_c \left(\frac{T_{\rm e}}{10^7 \; \rm{K}} \right) \left(\frac{r_c}{0.5h^{-1} \; {\rm Mpc}} \right) = 2.7 \pm 0.5.
\eeq
\item We apply the same analysis to the BAHAMAS suite of cosmological hydrodynamic simulations \citep{McCarthy2017}.  The results are marginally consistent, but the simulations predict a slightly lower mean tSZ signal of $\Delta y = (0.84 \pm 0.24) \times 10^{-8}$.
\item Our result for the over-density in filaments is compatible with the results of \cite{Epps2017}. They study the weak lensing signal between SDSS-III/BOSS LRG's and estimate $\delta \sim 4$, assuming a uniform-density, cylindrical filament model. 
After this paper was submitted, we learned of a similar analysis \citep{Graaff2017} which reaches conclusions that are consistent with ours.
\end{itemize}

Our investigation can be extended with larger spectroscopic surveys such as extended BOSS (eBOSS) in SDSS-IV and the Dark Energy Spectroscopic Instrument (DESI).  Their larger samples would improve the signal-to-noise and allow for a more detailed study of the physical conditions as a function of LRG properties, such as stellar mass and redshift. Large-area experiments with higher tSZ angular resolution, such as the Atacama Cosmology Telescope and the South Pole Telescope, would also help to ascertain the state of filament gas. \\

%%%%%%%%%%%%%%%%% APPENDICES %%%%%%%%%%%%%%%%%%%%%

\appendix

\section{Noise estimate from null $y$ map}
\begin{figure}
\includegraphics[width=\linewidth]{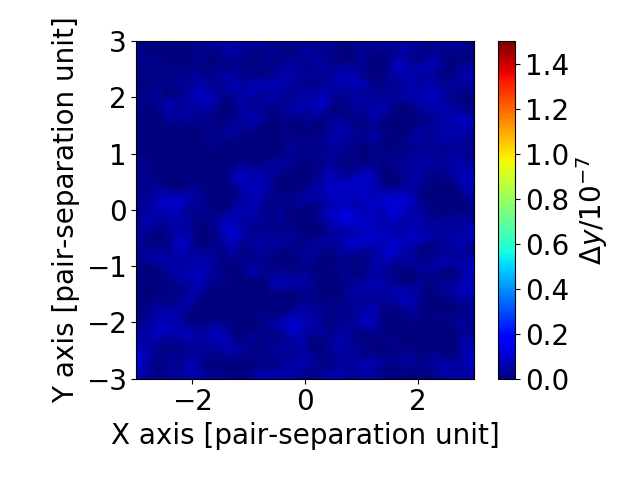}
\includegraphics[width=\linewidth]{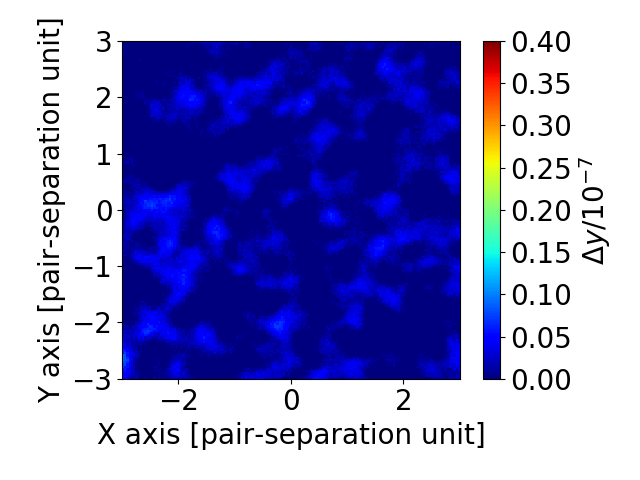}
\caption{{\it Top}: The null-$y$ map, constructed by subtracting the first half with the last half of MILCA $y$ map, stacked against 262,864 LRG pairs. {\it Bottom}: The null-$y$ map after subtracting circular halos.}
\label{f12}
\end{figure}

We estimate a noise contribution in the filament region using the null $y$ map constructed by subtracting the first half of MILCA $y$ map with the last half. This map nulls the tSZ signal as well as systematics such as foreground residuals in the $y$ map. For the null $y$ map, we perform the same analysis. As expected, we do not see any structure in the stacked map on the top panel of Figure~\ref{f12}. After the circular halo subtraction, we obtain $\Delta y = -0.03 \times 10^{-8}$ from the bridge region. The result suggests that the noise contribution is small compared to the mean filament $y$ signal.

\section{Halo mass distribution of DR12 LRGs}
\begin{figure}
\includegraphics[width=\linewidth]{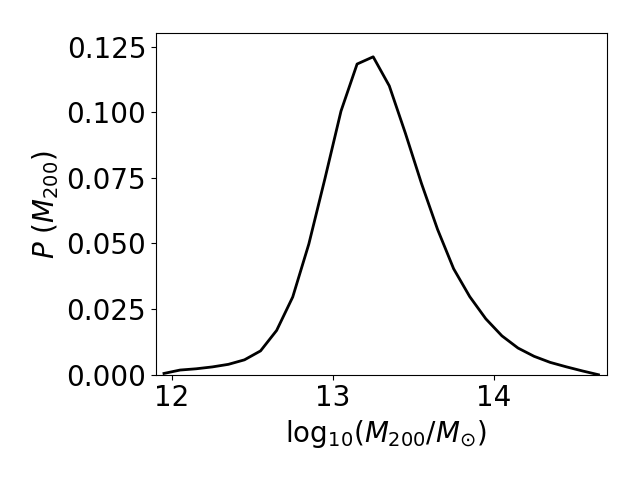}
\caption{The halo mass distribution of DR12 LRGs peaks at $M_{200} \sim 10^{13.3} M_{\odot}$.}
\label{f13}
\end{figure}

In \S\ref{sec:unbound}, we simulate a model $y$ map using only bound gas in SDSS DR12 LRGs with $10^{10} M_{\odot} < M_{\ast} < 10^{12} M_{\odot}$ and $0. < z < 0.8$. We select ``central`` LRGs described in \S\ref{sec:lrg}, which leaves $\sim$1,100,000 LRGs, and estimate halo masses of the LRGs with the stellar-to-halo mass relation of \cite{Coupon2015}, estimated in the CFHTLenS/VIPERS field by combining deep observations from near-UV to near-IR, supplemented by $\sim$70,000 secure spectroscopic redshifts, and analyzing galaxy clustering, galaxy-galaxy lensing and stellar mass function. The halo mass distribution is shown in Figure~\ref{f13}.

%%%%%%%%%%%%%%%%%%%%%%%%%%%%%%%%%%%%%%%%%%%%%%%%%%

\section*{Acknowledgement}
This research is funded by Canada's NSERC and CIFAR.

The data is based on observations obtained with Planck (http://www.esa.int/Planck), an ESA science mission with instruments and contributions directly funded by ESA Member States, NASA, and Canada, and observations obtained with SDSS-III, managed by the Astrophysical Research Consortium for the Participating Institutions of the SDSS-III Collaboration (http://www.sdss3.org/) funded by the Alfred P.Sloan Foundation, the Participating Institutions, the National Science Foundation, and the U.S. Department of Energy Office of Science. 

The authors thank Joop Schaye for his contributions to the BAHAMAS project. This work used the DiRAC Data Centric system at Durham University, operated by the Institute for Computational Cosmology on behalf of the STFC DiRAC HPC Facility (www.dirac.ac.uk<http://www.dirac.ac.uk>). This equipment was funded by BIS National E-infrastructure capital grant ST/K00042X/1, STFC capital grants ST/H008519/1 and ST/K00087X/1, STFC DiRAC Operations grant ST/K003267/1 and Durham University. DiRAC is part of the National E-Infrastructure.

This research has been also supported by funds from the European Research Council (ERC) under the European Union's Horizon 2020 research and innovation programme grant agreement ERC-2015-AdG 695561 (ByoPiC). The authors acknowledge fruitful discussions with the members of the ByoPiC project (https://byopic.eu/team).

AM has received funding from the European Union's Horizon 2020 research and innovation programme under the Marie Sklodowska-Curie grant agreement No. 702971.

Finally, we thank the reviewer for his/her thorough review and highly appreciate the comments and suggestions, which significantly contributed to improving the quality of the publication. 

%%%%%%%%%%%%%%%%%%%%%%%%%%%%%%%%%%%%%%%%%%%%%%%%%%


\footnotesize{
\setlength{\bibhang}{2.0em}
\setlength\labelwidth{0.0em}
\bibliographystyle{mnras}
\bibliography{bib}
}

%\bibliographystyle{mnras}
%\bibliography{bib} % if your bibtex file is called example.bib

%\begin{thebibliography}{99}
%\end{thebibliography}

%%%%%%%%%%%%%%%%%%%%%%%%%%%%%%%%%%%%%%%%%%%%%%%%%%

%%%%%%%%%%%%%%%%% APPENDICES %%%%%%%%%%%%%%%%%%%%%

%\appendix



%%%%%%%%%%%%%%%%%%%%%%%%%%%%%%%%%%%%%%%%%%%%%%%%%%

% Don't change these lines
\bsp	% typesetting comment
\label{lastpage}
\end{document}